\documentclass{aastex63}
\usepackage[caption=false]{subfig}
\usepackage{graphicx}
\usepackage{multirow}
\usepackage{graphicx}
\usepackage{booktabs}
\usepackage{amsmath,amssymb}
\usepackage[T1]{fontenc}
\usepackage{epstopdf}
\graphicspath{{./}{figures/}}

\begin{document}

\title{GeV $\gamma$-ray Emission of Compact Steep-Spectrum Source 4C +39.23B}

\correspondingauthor{Jin Zhang}
\email{j.zhang@bit.edu.cn}

\author{Ying Gu}
\affiliation{Guangxi Key Laboratory for Relativistic Astrophysics, School of Physical Science and Technology, \\
Guangxi University, Nanning 530004, People's Republic of China}

\author{Hai-Ming Zhang}
\affiliation{School of Astronomy and Space Science, Nanjing University, \\
Nanjing 210023, People's Republic of China}

\author{Ying-Ying Gan}
\affiliation{School of Physics, Beijing Institute of Technology, Beijing 100081, \\
People's Republic of China; j.zhang@bit.edu.cn}

\author{Jin Zhang\dag}
\affiliation{School of Physics, Beijing Institute of Technology, Beijing 100081, \\
People's Republic of China; j.zhang@bit.edu.cn}

\author{Xiao-Na Sun}
\affiliation{Guangxi Key Laboratory for Relativistic Astrophysics, School of Physical Science and Technology, \\
Guangxi University, Nanning 530004, People's Republic of China}

\author{En-Wei Liang}
\affiliation{Guangxi Key Laboratory for Relativistic Astrophysics, School of Physical Science and Technology, \\
Guangxi University, Nanning 530004, People's Republic of China}

\begin{abstract}
Thirteen yr observation data of 4FGL J0824.9+3915 with the Large Area Telescope on board the \emph{Fermi} Gamma Ray Space Telescope (\emph{Fermi}/LAT) are analyzed for revisiting whether 4C +39.23B, a compact steep-spectrum (CSS) source closed to a flat-spectrum radio quasar (FSRQ) 4C +39.23A in the $\gamma$-ray emitting region of 4FGL J0824.9+3915, is a $\gamma$-ray emitter. We find that the time-integrated $\gamma$-ray emission of 4FGL J0824.9+3915 is overwhelmingly dominated by 4C +39.23A. It shows significant variability at a 6.7$\sigma$ confidence level and the average $\gamma$-ray flux in the 0.1--300 GeV energy band is $(1.60\pm 0.15)\times10^{-8}$ ph cm$^{-2}$ s$^{-1}$ with a power-law photon spectral index of $2.48\pm0.05$. During MJD 57500--58500, 4FGL J0824.9+3915 is in a low state with a steady $\gamma$-ray flux. Analyzed the \emph{Fermi}/LAT observation data in this time interval, it is found that the TS values of the $\gamma$-ray emission from 4C +39.23A and 4C +39.23B are $\sim5$ and $\sim 31$, respectively, indicating that the $\gamma$-ray emission in this time interval is dominated by the CSS 4C +39.23B. The derived average flux in this time interval for 4C +39.23B is $(9.40\pm4.10)\times 10^{-9}$ ph cm $^{-2}$ s$^{-1}$ with $\Gamma_{\gamma}=2.45\pm0.17$. Attributing the spectral energy distribution (SED) of 4C +39.23B to the radiations from its core and extended region, we show that the SED can be represented with a two-zone leptonic model. Its $\gamma$-ray emission is contributed by the core region. The derived magnetic field strength and Doppler boosting factor of the core are 0.13 G and 6.5. Comparing 4C +39.23B with other $\gamma$-emitting CSSs and compact symmetric objects (CSOs) in the $\Gamma_{\gamma}-L_{\gamma}$ plane, it resembles CSSs.
 \end{abstract}


\section{Introduction}
\label{sec:intro}

As a young active galactic nucleus (AGN) population, compact steep-spectrum sources (CSSs) make up significant fraction ($\sim30\%$) of centimeter-wavelength-selected radio sources (O'Dea \& Saikia 2021 for a review). They are characterized by a steep radio spectrum ($\alpha\geq0.5$, $F_{\nu}\propto\nu^{-\alpha}$) and a compact dimension ($\leq15$ kpc, angular size $\leq1\arcsec \sim 2\arcsec$) without considering the projection effect (O'Dea 1998 for a review). More than 70\% CSSs may be intrinsically small rather than a projection effect (Fanti et al. 1990). It was proposed that CSSs may be the parent population of radio-loud narrow-line Seyfert 1 galaxies (RL-NLS1s, Caccianiga et al. 2014; Berton et al. 2016; Liao \& Gu 2020) because both CSSs and RL-NLS1s have similar radio properties (Komossa et al. 2006; Caccianiga et al. 2014; Gu et al. 2015), black hole (BH) masses, and Eddington ratios (Wu 2009). However, the link between RL-NLS1s and the young radio sources is also debated, for example, large linear sizes or the lack of absorption in the X-ray spectrum are observed for some $\gamma$-ray emitting RL-NLS1s (e.g., Orienti et al. 2015). Superluminal motion was observed in a few CSSs (Cotton et al. 1997; Taylor et al. 1995; Gawro\'{n}ski \& Kus 2006). Their kinematic ages estimated with the separation velocity of the hotspots and their radiative ages derived from the synchrotron radiation spectral break are roughly consistent within the error bars, indicating that CSSs are likely of young objects. A fraction of CSSs may eventually evolve into typical Fanaroff--Riley (FR) I or II radio galaxies (RGs; O'Dea 1998; Polatidis \& Conway 2003; Randall et al. 2011).

The $\gamma$-ray emitting CSSs are good candidates for investigating the jet properties of young AGNs. Only several CSSs so far are detected in the $\gamma$-ray band with the Large Area Telescope on board the \emph{Fermi} Gamma Ray Space Telescope (\emph{Fermi}/LAT, Abdollahi et al. 2020; Zhang et al. 2020). They share similar jet radiation properties to the other kinds of $\gamma$-ray emitting AGNs (Zhang et al. 2020). Recently, Principe et al. (2021) performed an analysis with the $\sim$11 yr \emph{Fermi}/LAT observation data to investigate the $\gamma$-ray emission of 162 young radio sources, and 11 sources among them have the significant $\gamma$-ray emission.

4C +39.23B is a CSS. It has a double morphology (component-E and component-W) with a total extent of $\sim$60 mas in the very long baseline array (VLBA) images at 5.0 and 8.4 GHz (Orienti et al. 2004). The two components have a steep-spectrum $\alpha\geq 1$ in the range from 1.7 to 5.0 GHz. Component-E is brighter than component-W, and its core is not detected. The association between 4C +39.23B and the $\gamma$-ray source 3FGL J0824.9+3916 is reported in the Third \emph{Fermi} LAT AGN catalog (3LAC, Ackermann et al. 2015, see also Massaro et al. 2015). However, it is controversial whether 4C +39.23B is a $\gamma$-ray emitter since a flat-spectrum radio quasar (FSRQ) 4C +39.23A ($z=1.21568$, P\^{a}ris et al. 2018) is also reported to be the counterpart of 3FGL J0824.9+3916 in the 3LAC (Ackermann et al. 2015). Massaro et al. (2015) also argued that 4C +39.23A may be responsible for the $\gamma$-ray detections of 3FGL J0824.9+3916. In the updated Fourth \emph{Fermi} LAT Source Catalog (4FGL, Abdollahi et al. 2020), 4FGL J0824.9+3915 is suggested to be associated with 4C +39.23A. We dedicate to investigate whether 4C +39.23B is a $\gamma$-ray emitter by analyzing the $\sim$13 yr \emph{Fermi}/LAT observational data of 4FGL J0824.9+3915. The photometric redshift of 4C +39.23B is $z\sim1.0-1.4$ (Lilly 1989) and $z=1.21$ is used in our analysis (see also Law-Green et al. 1995; Roche et al. 1998). Throughout, $H_{0}$=71 km s$^{-1}$ Mpc$^{-1}$, $\Omega_{m}$=0.27, and $\Omega_{\Lambda}$=0.73 are used.

\section{\emph{Fermi}/LAT Data Analysis}

The 13 yr (covering from 2008 August 4 to 2021 October 31) \emph{Fermi}/LAT observational data are adopted for our analysis. The publicly available software \textit{Fermitools} (ver.\ 1.2.23) with the binned likelihood analysis method are used to analyze the data. The region of interest (ROI) is selected as a $14\degr \times 14\degr$ square region centered at 4FGL J0824.9+3915 (R.A.=126.24$\degr$, Decl.=39.26$\degr$). We select only the $\gamma$-ray events in the $0.1-300$ GeV energy range with the standard data quality selection criteria ``$(DATA\_QUAL > 0) \&\& (LAT\_CONFIG == 1)$''. Considering the contamination of the $\gamma$-ray Earth limb, we set the maximum zenith angle to be 90$\degr$. The instrument response function of $P8R3\_SOURCE\_V2$ is used.

We first examine the possible new sources in the background. We set a model of known $\gamma$-ray sources in the ROI. The model include the isotropic emission (``iso\_P8R3\_V3\_v1.txt''), the diffuse Galactic interstellar emission (IEM, ``$gll\_iem\_v07.fits$''), and all $\gamma$-ray point sources and extended sources listed in the \emph{Fermi} LAT Fourth Source Catalog Data Release 2 (4FGL-DR2, Abdollahi et al. 2020; Ballet et al. 2020). The parameters of the spectral models of the $\gamma$-ray point sources within ROI and the IEM and isotropic emission models are kept free. The maximum likelihood test statistic (TS) is used to estimate the significance of $\gamma$-ray signals, which is defined by TS$=2(\ln\mathcal{L}_{1}-\ln\mathcal{L}_{0})$, where $\mathcal{L}_{1}$ is the likelihood value for the background including the point source and $\mathcal{L}_{0}$ is the likelihood value for the background without the point source (null hypothesis). Using the package \emph{gtfindsrc}, we find a new $\gamma$-ray source located at (R.A.=127.36$\degr$$\pm0.04\degr$, Decl.=40.33$\degr$$\pm0.04\degr$) with TS$\sim$45.8, which is 1.55$\degr$ away from 4FGL J0824.9+3915. The average flux of this new source is $(2.60\pm1.00)\times 10^{-9}$ ph cm$^{-2}$ s$^{-1}$ with a photon spectral index of 2.28$\pm0.15$ in the 0.1--300 GeV energy band.

\begin{figure}
 \centering
   \includegraphics[angle=0,scale=0.49]{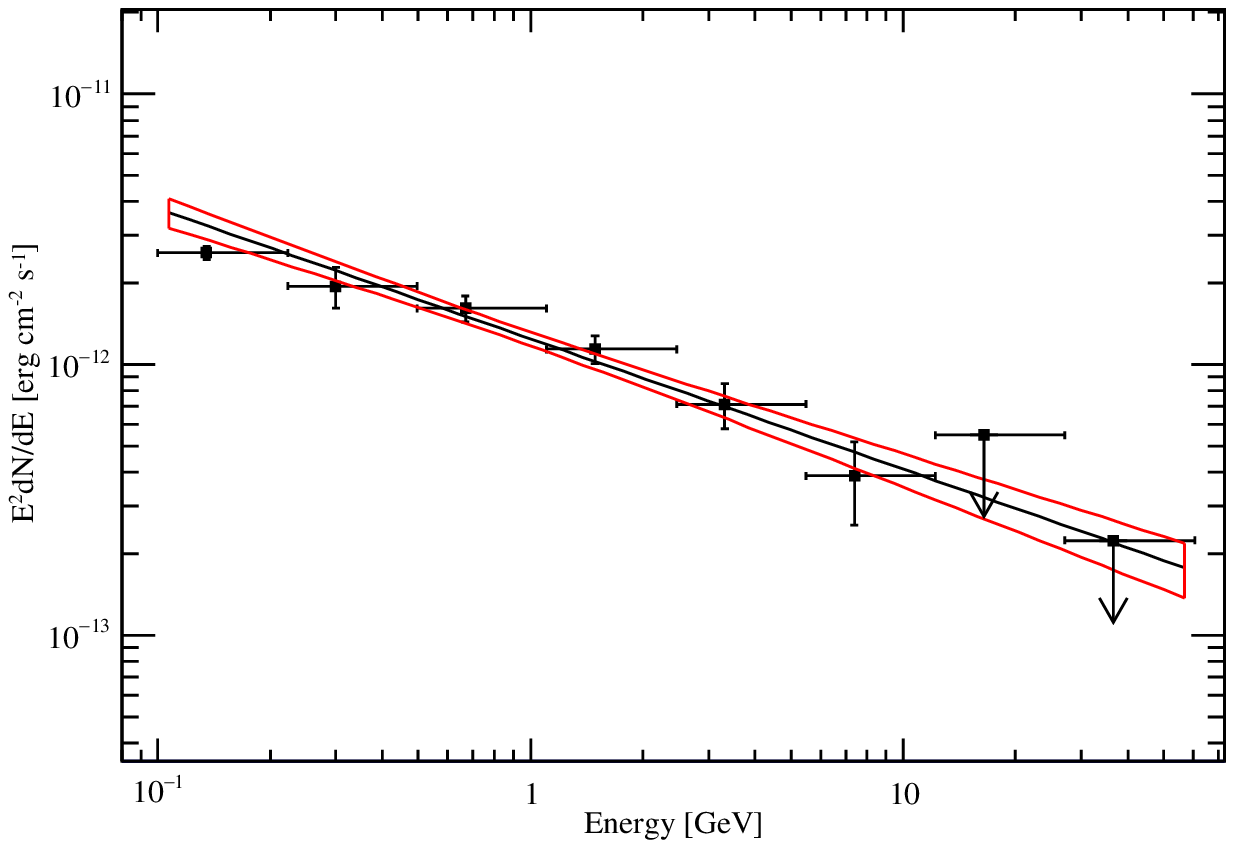}
   \includegraphics[angle=0,scale=0.27]{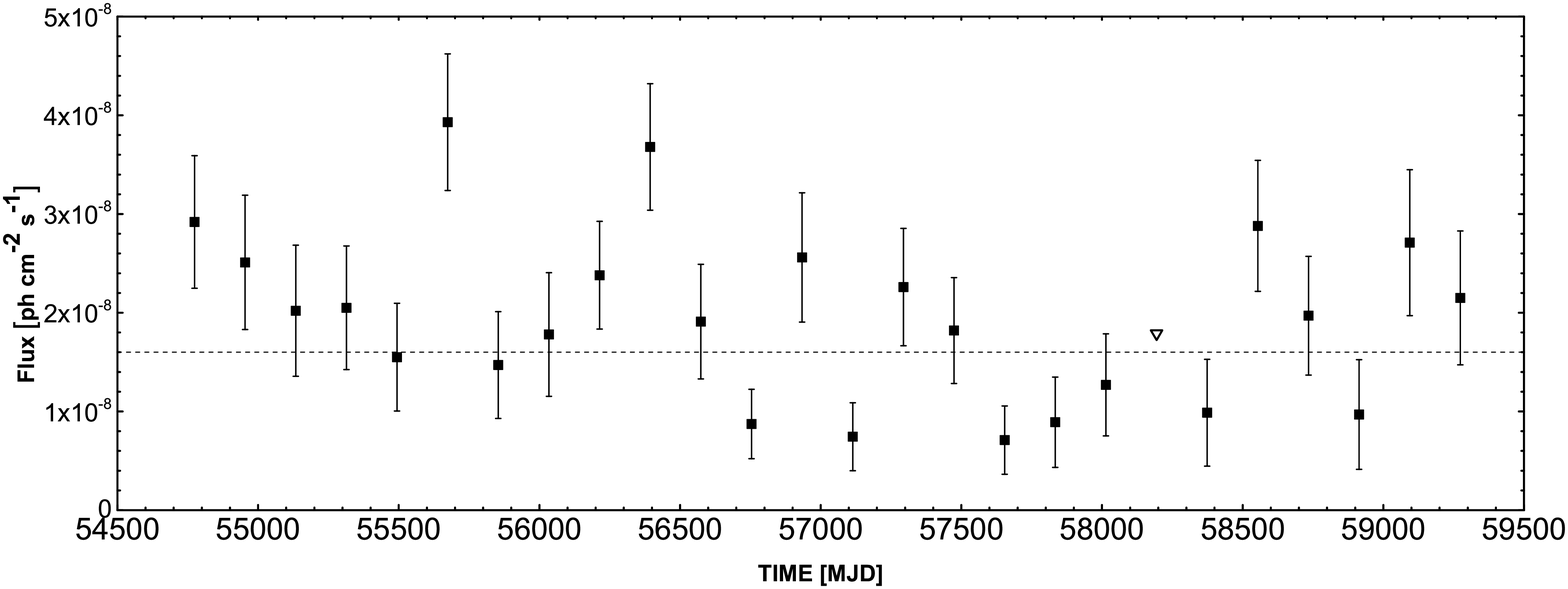}
\caption{\emph{Left panel}: the $\sim$13 yr average energy spectrum along the power-law function fit for 4FGL J0824.9+3915. \emph{Right panel}: the long-term light curve for 4FGL J0824.9+3915 with time bins of 180 days. The inverted triangle indicates TS$<$9 for this time bin and the horizontal dashed line represents the $\sim$13 yr average flux of 4FGL J0824.9+3915 observed by the \emph{Fermi}/LAT.}\label{LAT}
\end{figure}

\begin{figure}
 \centering
   \includegraphics[angle=0,scale=0.25]{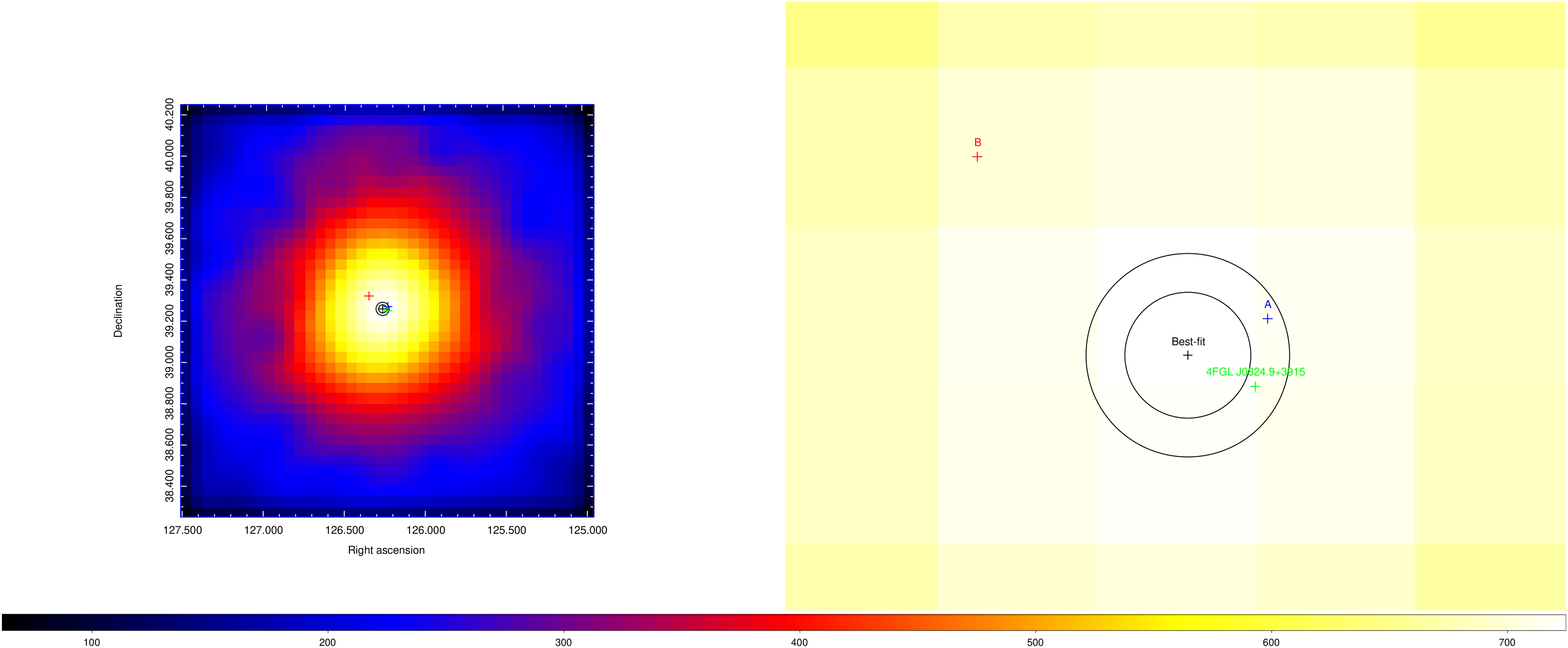}
\caption{\emph{Left panel}: 2.0$\degr\times2.0\degr$ TS map in the 0.1--300 GeV energy band observed by the \emph{Fermi}/LAT covering MJD 54682--59518 for 4FGL J0824.9+3915 (green cross). The black symbols indicate the best-fit position (black cross) of the $\sim$13 yr \emph{Fermi}/LAT observation data together with the 68\% and 95\% error circles (black circles). The blue and red crosses show the radio positions of 4C +39.23A (A) and 4C +39.23B (B), respectively. \emph{Right panel}: zoom in the central region of the left panel. The TS map is created with a pixel size of 0.05$\degr$ and smoothed with a Gaussian kernel of 0.3$\degr$.}\label{TSmap_A}
\end{figure}

We add the new point source as a background source to our source model and perform the analysis again. We use TS$_{\rm curv}=2\log(\mathcal{L}_{\rm log-normal}/\mathcal{L}_{\rm power-law})$ to evaluate the curvature significance of a spectrum. A spectrum is regarded as significantly curved if its $\rm TS_{curv}>9$ ($\rm TS_{curv}=9$ corresponding to 3$\sigma$, Abdollahi et al. 2020). We find TS$_{\rm curv}\sim4$ for 4FGL J0824.9+3915, indicating that the spectral shape of 4FGL J0824.9+3915 is well described by a power-law function. We thus adopt a power-law spectral function to describe the spectrum of 4FGL J0824.9+3915 (Figure \ref{LAT}), i.e., $dN(E)/dE=N_0(E/E_0)^{-\Gamma_{\gamma}}$, where $\Gamma_{\gamma}$ is the photon spectral index and $N(E)$ is the photon distribution as a function of energy. The TS map of 4FGL J0824.9+3915 derived with the $\sim$13 yr \emph{Fermi}/LAT observation data is shown in Figure \ref{TSmap_A}. Using the package \emph{gtfindsrc}, we obtain the best-fit position of 4FGL J0824.9+3915, i.e., (R.A.=126.26\degr$\pm0.02\degr$, Decl.=39.27\degr$\pm0.02\degr$). The position of the $\gamma$-ray source reported in the 4FGL-DR2 (Abdollahi et al. 2020; Ballet et al. 2020) is within the 95\% error circle of our analysis result. The TS value of the source is 492.7. The average flux of 4FGL J0824.9+3915 is $(1.60\pm0.15)\times 10^{-8}$ ph cm$^{-2}$ s$^{-1}$ with $\Gamma_{\gamma}=2.48\pm0.05$ in the 0.1--300 GeV energy band.

Figure \ref{LAT} shows the the light curve of 4FGL J0824.9+3915 in a time bin of 180 days with  a criterion of TS$\geq$9 (approximately corresponds to 3$\sigma$ detection, Mattox et al. 1996). Significant variability is observed. The variability index TS$_{\rm var}$ is commonly used to estimate the significance of variability for a source, which is defined as
\begin{equation}
\rm TS_{\rm var}=2\sum_{i=1}^N [ log(\mathcal{L}_{i}(F_i)) - log(\mathcal{L}_{i}(F_{\rm glob}))],
\end{equation}
where $F_i$ is the fitting flux for the bin $i$, $\mathcal{L}_{i}(F_i)$ is the likelihood corresponding to the bin $i$, and $F_{\rm glob}$ is the best fit flux for the glob time by assuming a constant flux (Nolan et al. 2012). $\rm TS_{\rm var}=51.7$ indicates that the variability is at $3\sigma$ confidence level in a $\chi^2_{N-1}$(TS$_{\rm var}$) distribution with $N-1=25$ degrees of freedom, $N$ is the number of time bins. We obtain TS$_{\rm var}$=104.3, which corresponds to a 6.7$\sigma$ confidence level, indicating that 4FGL J0824.9+3915 is a variable source\footnote{The average flux of 4FGL J0824.9+3915 reported in the 4FGL-DR2 is $F_{1-100\ \rm GeV}=(6.13\pm0.45)\times10^{-10}$ ph cm$^{-2}$ s$^{-1}$ with $\Gamma_{\gamma}=2.43\pm0.05$ and the variability index is TS$_{\rm var}=35.7$ (the bin size of light curve is one year, Abdollahi et al. 2020; Ballet et al. 2020) corresponding to a 3.7$\sigma$ confidence level. With the 13 yr \emph{Fermi}/LAT data, we obtain $F_{1-100\ \rm GeV}=(5.40\pm0.50)\times 10^{-10}$ ph cm$^{-2}$ s$^{-1}$ with $\Gamma_{\gamma}=2.48\pm0.05$ and TS$_{\rm var}=104.3$ corresponding to a 6.7$\sigma$ confidence level for 4FGL J0824.9+3915. Our results are consistent with that reported in the 4FGL-DR2.}.

\section{Association between 4FGL J0824.9+3915 and 4C +39.23A/B }
We show the radio positions of 4C +39.23A (R.A.=126.23$\degr$, Decl.=39.28$\degr$) and 4C +39.23B (R.A.=126.35$\degr$, Decl.=39.33$\degr$) together with the 68\% and 95\% error circles of the best-fit position for 4FGL J0824.9+3915 in Figure \ref{TSmap_A}. The radio position of 4C +39.23A falls into the 95\% error circle of the best-fit position, indicating that 4FGL J0824.9+3915 may be spatially associated with 4C +39.23A. 4C +39.23B is located at 0.12$\degr$ away from the best-fit position and outside of the 95\% error circle. Note that the angular resolution of the \emph{Fermi}/LAT increases with the energy band from $<3^{\circ}$.5 at 100 MeV to $<0^{\circ}$.15 at $>$10 GeV (Atwood et al. 2009). Using the \textit{gtsrcprob} tool, we find that the maximum energy of the detection photons for 4FGL J0824.9+3915 is only 16.9 GeV. Considering the low spacial resolution of the \emph{Fermi}/LAT, one still cannot convincingly exclude the contribution from 4C +39.23B to the $\gamma$-ray flux of 4FGL J0824.9+3915. It is also possible that there is another unidentified source nearby the position of 4FGL J0824.9+3915. Thus, the $\gamma$-ray emission of 4FGL J0824.9+3915 may be contributed from 4C +39.23A, 4C +39.23B, and/or the unidentified source. We further check this issue in the following scenarios (see also in Table 1).

\begin{itemize}
 \item Scenario I: The $\gamma$-ray detections are fully attributed to an unidentified source at the position of 4FGL J0824.9+3915. In this scenario, we obtain TS=492.7, $F_{0.1-300\ \rm GeV}=(1.60\pm0.15)\times10^{-8}$ ph cm$^{-2}$ s$^{-1}$, and $\Gamma_\gamma=2.48\pm0.05$.

 \item Scenario II: The $\gamma$-ray detections of 4FGL J0824.9+3915 are assumed to be fully attributed to 4C +39.23A by setting the $\gamma$-ray position at 4C +39.23A. In this scenario, we obtain TS=491.2, $F_{0.1-300\ \rm GeV}=(1.60\pm0.16)\times10^{-8}$ ph cm$^{-2}$ s$^{-1}$, and $\Gamma_\gamma=2.48\pm0.05$.

 \item Scenario III: The $\gamma$-ray detections of 4FGL J0824.9+3915 are assumed to be fully attributed to 4C +39.23B by setting the $\gamma$-ray position at 4C +39.23B. In this scenario, we obtain TS=458.6, $F_{0.1-300\ \rm GeV}=(1.60\pm0.16)\times10^{-8}$ ph cm$^{-2}$ s$^{-1}$, and $\Gamma_\gamma=2.50\pm0.05$.

 \item Scenario IV: The $\gamma$-ray detections of 4FGL J0824.9+3915 are attributed to both 4C +39.23A and 4C +39.23B. In this scenario, we get TS=316.1, $F_{0.1-300\ \rm GeV}=(1.30\pm0.34)\times10^{-8}$ ph cm$^{-2}$ s$^{-1}$, and $\Gamma_\gamma=2.48\pm0.07$ for 4C +39.23A, and TS=7.8, $F_{0.1-300\ \rm GeV}<0.59\times10^{-8}$ ph cm$^{-2}$ s$^{-1}$ (an upper-limit is presented if TS$<$9), and $\Gamma_\gamma=2.52\pm0.28$ for 4C +39.23B, respectively.

 \item Scenario V: The $\gamma$-ray detections are attributed to 4C +39.23A, 4C +39.23B, and an unidentified source at the position of 4FGL J0824.9+3915. In this scenario, the fits are not convergent.
 \end{itemize}

One can find that the results from the scenarios I--IV are consistent, indicating that the $\gamma$-rays of 4FGL J0824.9+3915 are dominated by 4C +39.23A and the hypothetically unidentified source in the position of 4FGL J0824.9+3915 should be 4C +39.23A. In the scenario V, if the $\gamma$-ray detections are attributed to 4C +39.23A, 4C +39.23B, and an unidentified source in the position of 4FGL J0824.9+3915, the fits are not convergent, indicating that this scenario is unreasonable. As shown in scenarios I--III, the derived fluxes and photon spectral indices are almost the same, and thus it is also possible that 4C +39.23B contributes to the $\gamma$-rays. Taking the contributions from both 4C +39.23A and 4C +39.23B into account, we obtain TS=316.1 for 4C +39.23A and TS=7.8 for 4C +39.23B, respectively. In this case, 4C +39.23A approximatively contributes to 81\% of the $\gamma$-ray flux. One can also find that the contribution from 4C +39.23B to the $\gamma$-rays is not negligible.

\begin{figure}
 \centering
 \includegraphics[angle=0,scale=0.25]{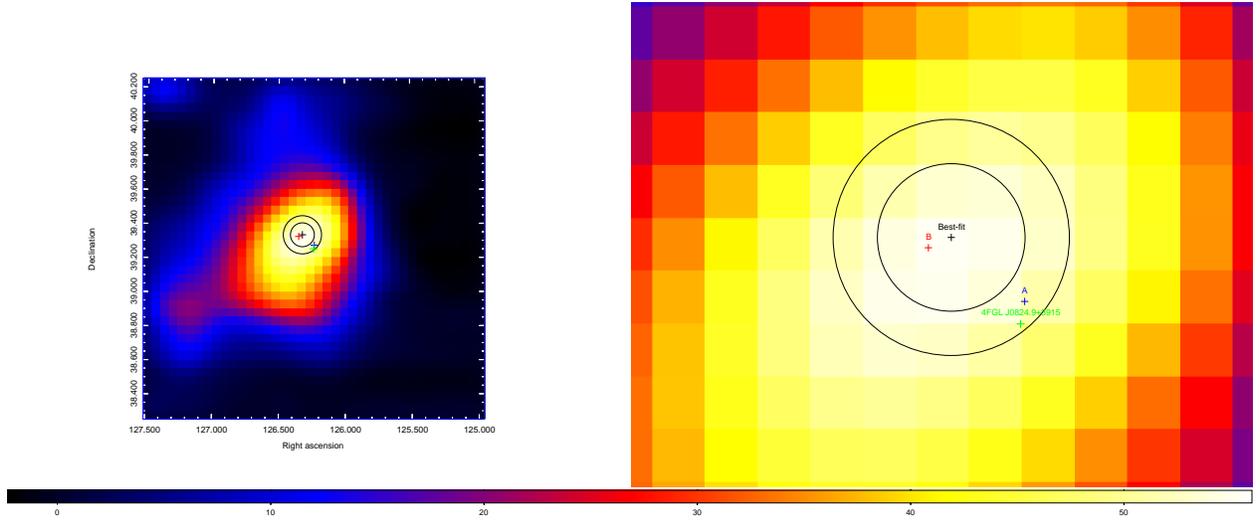}
\caption{\emph{Left panel}: 2.0$\degr$ $\times$ 2.0$\degr$ TS map in the 0.1--300 GeV energy band observed by the \emph{Fermi}/LAT covering MJD 57500--58500 for 4FGL J0824.9+3915 (green cross). The black symbols indicate the best-fit position (black cross) of the \emph{Fermi}/LAT observation data (covering MJD 57500--58500) together with the 68\% and 95\% error circles (black circles). The blue and red crosses show the radio positions of 4C +39.23A (A) and 4C +39.23B (B), respectively. \emph{Right panel}: zoom in the central region of the left panel. The TS map is created with a pixel size of 0.05$\degr$ and smoothed with a Gaussian kernel of 0.3$\degr$.}\label{TSmap_B}
\end{figure}

\begin{figure}
 \centering
   \includegraphics[angle=0,scale=0.64]{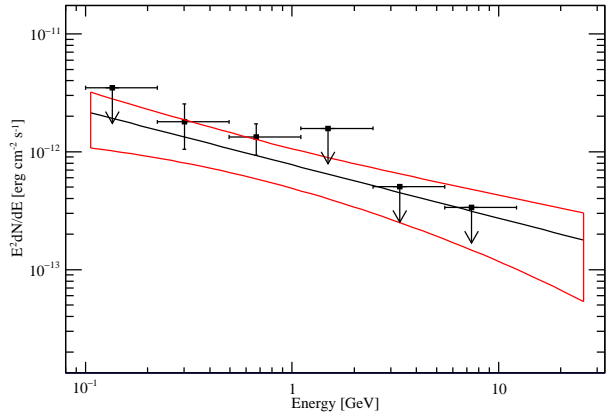}
\caption{The average energy spectrum observed by the \emph{Fermi}/LAT covering MJD 57500--58500 for 4C +39.23B. The black solid line represents the fitting result with a power-law function and the red bow shows the 1$\sigma$ error. If TS$<$9, an upper-limit is presented for this energy bin. }\label{LATSED_B}
\end{figure}

Note that 4C +39.23A is a FSRQ, a sub-class of blazar, which is characterized by violent variability. The above analysis shows that 4FGL J0824.9+3915 is a variable source with a 6.7$\sigma$ confidence level. These results convincingly indicate that the detected $\gamma$-rays are overwhelmingly dominated by 4C +39.23A. The weak $\gamma$-ray emission from 4C +39.23B is of our interest. As shown in Figure \ref{LAT}, 4FGL J0824.9+3915 is in the low state from MJD 57500 to MJD 58500, which would be offer an opportunity for estimating the contribution of 4C +39.23B to the $\gamma$-rays of 4FGL J0824.9+3915. We thus reanalyze the \emph{Fermi}/LAT observation data in this time interval. The new source is removed in our analysis. The derived TS map is given in Figure \ref{TSmap_B}. The \textit{gtfindsrc} best-fit position is located at (R.A.=126.32$\degr$$\pm0.07\degr$, Decl.=39.34$\degr$$\pm0.07\degr$) with TS=52.3 and an average flux of $(1.10\pm0.30)\times10^{-8}$ ph cm$^{-2}$ s$^{-1}$. 4C +39.23B is closer to the best-fit position of the $\gamma$-ray source than 4C +39.23A and falls into the 68\% error circle of the best-fit position. The position of 4C +39.23A is out of the 68\% error circle, but is still within the 95\% error circle. Again, assuming that both 4C +39.23A and 4C +39.23B are responsible for the $\gamma$-ray detections in this time interval, we obtain TS=31.0 ($>5\sigma$) with an average flux of $(9.40\pm4.10)\times10^{-9}$ ph cm$^{-2}$ s$^{-1}$ for 4C +39.23B while only TS=4.8 is derived for 4C +39.23A. We estimate the probability of association between the $\gamma$-ray source and 4C +39.23B in this time interval with a VLBA catalog\footnote{To further check our result, we also estimate the probability of association between the $\gamma$-ray source and 4C +39.23A using the $\sim$13 yr \emph{Fermi}/LAT data and obtain 99.1\% while 99.85\% was reported in the 4FGL-DR2 (Abdollahi et al. 2020; Ballet et al. 2020). }, where a Bayesian method with the \textit{gtsrcid} tool (Abdo et al. 2010) is used, and the derived association probability is 98.6\%. Hence, the $\gamma$-rays of 4FGL J0824.9+3915 in this time interval are dominated by 4C +39.23B. The derived average energy spectrum covering MJD 57500--58500 for 4C +39.23B is shown in Figure \ref{LATSED_B}, which is fitted by a power-law model with $\Gamma_{\gamma}=2.45\pm0.17$.

\section{Jet Properties derived from Broadband SED Modeling}

\begin{figure}
 \centering
   \includegraphics[angle=0,scale=0.35]{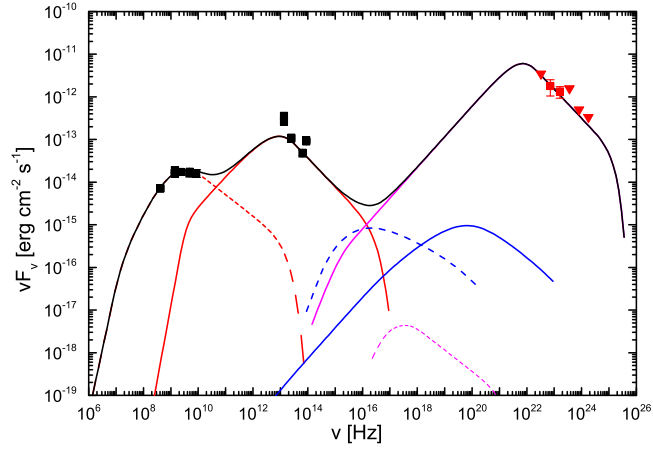}
\caption{Observed SED along with the two-zone leptonic model fitting curves. The data marked as black solid symbols are taken from the NED and ASDC. The red solid symbols in the $\gamma$-ray band indicate the average energy spectrum of the \emph{Fermi}/LAT observations for 4C +39.23B (same as in Figure \ref{LATSED_B}), where the down-triangles represent upper-limits. The thick black solid line is the sum of emission from each component: synchrotron radiation (red lines), SSC process (blue lines), and EC process (magenta lines), and among them the solid lines and dashed lines represent the radiations from the core and extended region, respectively.}\label{SED}
\end{figure}

As discussed above, 4C +39.23B would be one more valuable member of the $\gamma$-ray emitting CSSs. Using the derived average energy spectrum of 4C +39.23B at the GeV energy band (Figure 4) and other multi-wavelength data from the NASA/IPAC Extragalactic Database (NED) and ASI Science Data Center (ASDC)\footnote{https://tools.ssdc.asi.it/SED/}, we establish the broadband spectral energy distribution (SED) of 4C +39.23B to investigate its multi-wavelength radiation properties, as illustrated in Figure \ref{SED}. We also use the two-zone leptonic model as reported in Zhang et al. (2020) to reproduce the broadband SED of 4C +39.23B. The synchrotron (syn), synchrotron-self-Compton (SSC), and external Compton (EC) scattering of the relativistic electrons in both the core and extended region are considered. The radio emission below 10 GHz in the SED should be dominated by the radiation of the extended region on account of the significant synchrotron-self-absorption effect on the radio emission of the core.

The electron distribution in both regions is assumed as a broken power-law, characterized by a density parameter $N_0$, a break energy $\gamma_{\rm b}$, and indices $p_1$ and $p_2$ in the range of [$\gamma_{\min}, \gamma_{\max}$]. The radiation region is taken as a homogeneous sphere with radius $R$. The predicted spectrum by the leptonic model is decided by the following parameters: the size of the radiating region $R$, the strength of
magnetic field $B$, and the Doppler boosting factor $\delta$ together with the six electron spectrum parameters, where $\delta=1/(\Gamma-\sqrt{\Gamma^2-1}\cos\theta)$, $\Gamma$ and $\theta$ are the bulk Lorenz factor and viewing angle of the radiation region. The Klein-Nishina effect and the absorption of high energy $\gamma$-ray photons by extragalactic background light (Franceschini et al. 2008) are also taken into account during SED modeling.

For the core region, we take $R=\delta c\Delta t/(1+z)\sim2.1\times10^{17}\delta$, where $\Delta t=180$ days, and thus we consider the photons from the torus provide the seed photons for the inverse Compton process (IC/torus) of the relativistic electrons in the core region. No viewing angle is available for 4C +39.23B, we thus assume $\delta$=$\Gamma$ during SED modeling. For the extended region, the size of the emitting region $R$ is estimated by the angular radius of 60 mas at the radio band (Orienti et al. 2004). We do not consider the relativistic effect of the extended region and assume $\Gamma=\delta=1$ during SED modeling. The IC process of the cosmic microwave background (IC/CMB) by the relativistic electrons in the extended region is considered. The syn+SSC+IC/CMB model under the equipartition condition ($U_{B}=U_{\rm e}$) is used to reproduce the radiation of the extended region, where $U_{B}$ and $U_{\rm e}$ are the energy densities of magnetic fields and electrons, respectively. The fitting result of the broadband SED for 4C +39.23B is given in Figure \ref{SED} and the derived modeling parameters are listed in Table 2.

As shown in Figure \ref{SED}, the SED of 4C +39.23B can be well explained by the model and the $\gamma$-rays are produced by the IC/tours process. Note that we cannot give a constraint on the fitting parameters due to the limited observation data and only obtain a set of model parameters by artificially adjusting the model parameters to make an acceptable fit. For comparison, the SED fitting parameters of other $\gamma$-ray emitting CSSs (taken from Zhang et al. 2020) are also given in Table 2, the derived fitting parameters of 4C +39.23B are consistent with them.

\section{Conclusions and Discussion}

We have reanalyzed the $\sim$13 yr \emph{Fermi}/LAT data of 4FGL J0824.9+3915 for evaluating the contributions from radio sources 4C +39.23A and 4C +39.23B to the $\gamma$-ray detections. We showed that 4FGL J0824.9+3915 is detected with a TS value of 492.7. The derived average $\gamma$-ray flux of 4FGL J0824.9+3915 is $(1.60\pm 0.15)\times 10^{-8}$ ph cm$^{-2}$ s$^{-1}$ with a photon spectral index of $2.48\pm0.05$ in the 0.1--300 GeV energy band. In the derived TS map, the radio position of 4C +39.23A falls into the $95\%$ error circle of the best-fit position, while the radio position of 4C +39.23B is away from the best-fit position and outside of the $95\%$ error circle. Considering the contributions of the two sources, we found that the TS values are 316.1 and $\sim 7.8$ for 4C +39.23A and 4C +39.23B, respectively, indicating that 4C +39.23A is dominant to the total $\gamma$-rays of 4FGL J0824.9+3915. The $\gamma$-ray emission of 4FGL J0824.9+3915 is significantly variable at a 6.7$\sigma$ confidence level, which is overwhelmingly dominated by the radiations of 4C +39.23A. During MJD 57500--58500, the $\gamma$-ray emission of 4FGL J0824.9+3915 is in a low state with a steady flux. To reanalyze the \emph{Fermi}/LAT observation data in this time interval, it was found that the TS values of the $\gamma$-ray emission from 4C +39.23A and 4C +39.23B are 4.8 and 31.0, respectively, indicating that the $\gamma$-ray emission in this time interval is dominated by the CSS 4C +39.23B. The derived probability of association between the $\gamma$-ray source and 4C +39.23B in this time interval with a Bayesian method is 98.6\%. The average flux of 4C +39.23B covering MJD 57500--58500 is $(9.40\pm4.10)\times 10^{-9}$ ph cm $^{-2}$ s$^{-1}$ with a photon spectral index of $2.45\pm 0.17$. Attributing the broadband SED of 4C +39.23B to the radiations from the core and extended region, we showed that the SED can be represented by a two-zone leptonic model. Its $\gamma$-ray emission is contributed by the core region, which has a magnetic field strength of 0.13 G and a Doppler boosting factor of 6.5.

\begin{figure}
 \centering
   \includegraphics[angle=0,scale=0.73]{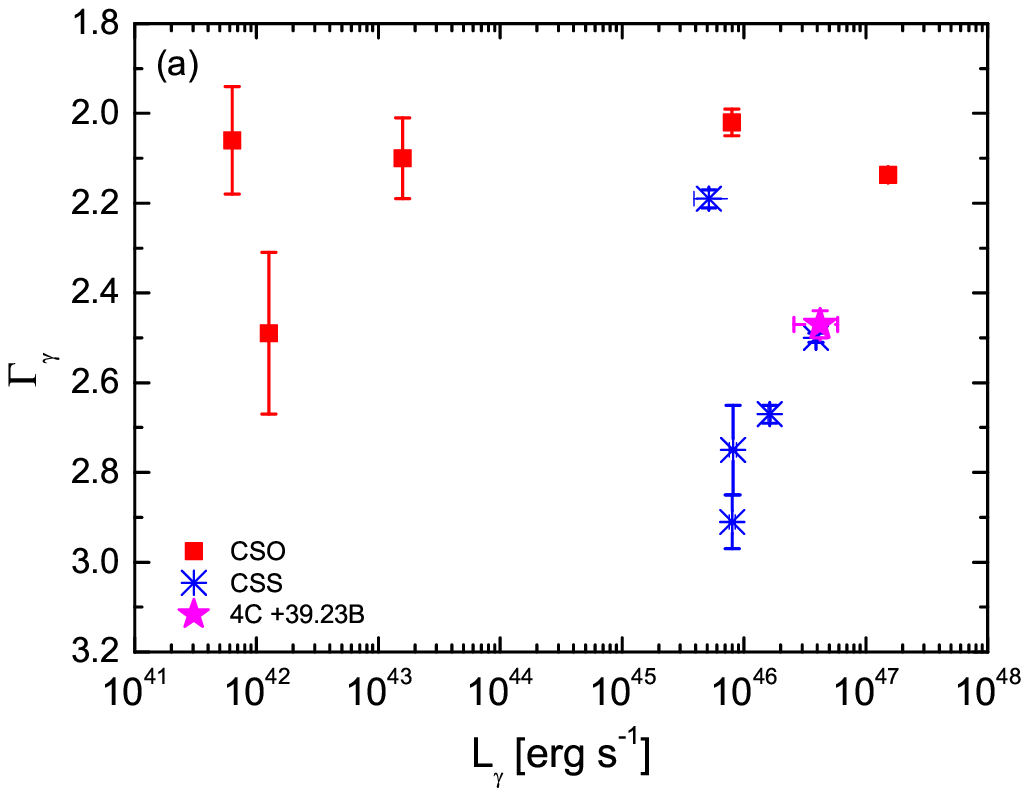}
   \includegraphics[angle=0,scale=0.73]{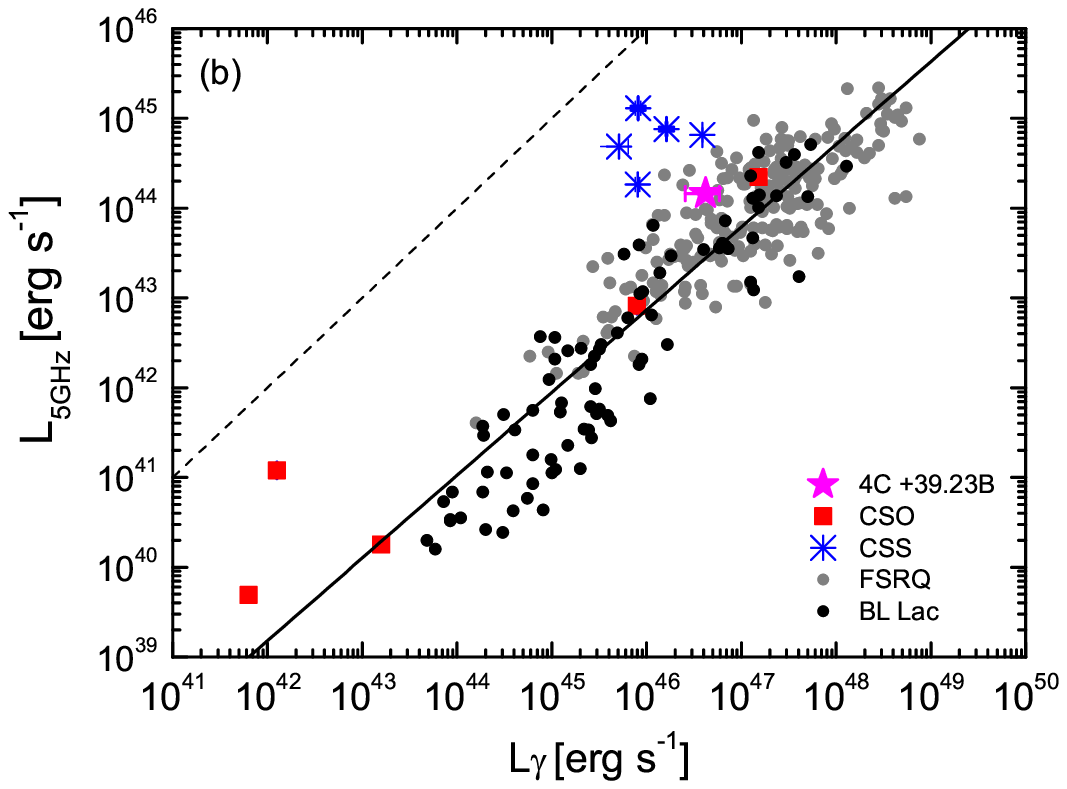}
\caption{\emph{Panel (a)}: Comparison of 4C +39.23B with other CSSs and CSOs in the $\Gamma_{\gamma}-L_{\gamma}$ plane. $\Gamma_{\gamma}$ and $L_{\gamma}$ of 4C +39.23B are derived with the \emph{Fermi}/LAT observation data covering MJD 57500--58500. $\Gamma_{\gamma}$ and $L_{\gamma}$ of the $\gamma$-ray emitting CSSs and CSOs are taken from Zhang et al. (2020), Abdollahi et al. (2020), and Gan et al. (2021), respectively. \emph{Panel (b)}: Comparison of 4C +39.23B with other CSSs and COSs as well as blazars in the $L_{\rm 5GHz}-L_{\gamma}$ plane. The dashed line is the equality line while the solid line is the linear fit in log scale for all the blazar data, i.e., $\log L_{\rm 5GHz}=(0.47\pm1.16)+(0.92\pm0.02)\log L_{\gamma}$. $L_{\rm 5GHz}$ of these $\gamma$-ray emitting CSSs and CSOs are taken from NED and Lister et al. (2020). $L_{\rm 5GHz}$ of 4C +39.23B are taken from the ASDC. The blazar data are taken from Fan \& Wu et al. (2018).}\label{gamma-L}
\end{figure}

Our results suggested that 4C +39.23B is a weak $\gamma$-ray emitter. The VLBA images of 4C +39.23B show a double morphology without detected core (Orienti et al. 2004) and the total extent of $\sim60$ mas corresponds to $\sim0.5$ kpc ($<$1 kpc). 4C +39.23B would be a compact symmetric object (CSO) considering these radio properties. We compare it with other $\gamma$-ray emitting young sources in the $\Gamma_\gamma-L_\gamma$ plane in Figure \ref{gamma-L}(a), where the five CSOs are CTD 135 (Gan et al. 2021), PMN J1603--4904 (M\"{u}ller et al. 2014), PKS 1718--649 (Migliori et al. 2016), NGC 3894 (Principe et al. 2020), and TXS 0128+554 (Lister et al. 2020), and the five CSSs include 3C 138, 3C 216, 3C 286, 3C 309.1, and 3C 380 (Zhang et al. 2020 and references therein)\footnote{We do not take 4C 15.05 into account here since its type is still debated.}. One can observe that 4C +39.23B almost has the highest $\gamma$-ray luminosity (lower than CTD 135, Gan et al. 2021). In the $\Gamma_{\gamma}-L_{\gamma}$ plane, the location of 4C +39.23B is consistent with other five $\gamma$-ray emitting CSSs, and almost overlaps with that of 3C 380. Comparing with the five $\gamma$-ray emitting CSOs, except for CTD 135, 4C +39.23B has the higher $\gamma$-ray luminosity than those CSOs, but has a steeper spectrum. As reported in Zhang et al. (2020), except for 3C 216, the other $\gamma$-ray emitting CSSs appear as variable sources and display obvious variability. For the five $\gamma$-ray emitting CSOs, both CTD 135 and PMN J1603--4904 display the obvious variability in $\gamma$-rays while the other three CSOs are not ( Principe et al. 2021; Lister et al. 2020). As illustrated in Figure \ref{gamma-L}(b), these young radio sources have the lower ratios of the $\gamma$-ray luminosity to the radio luminosity ($L_{\rm 5GHz}$) than blazars. The $\gamma$-rays of blazars are totally dominated by the radiations of their core jets, and thus the $\gamma$-rays are strongly amplified due to the Doppler boosting effect while the radio radiations are mainly from the more extended regions. Comparing with blazars, these young radio sources have the weakly relativistic effect and thus the lower ratios of $L_{\gamma}$ to $L_{\rm 5GHz}$.

On the basis of the above analysis results, the $\gamma$-rays of 4FGL J0824.9+3915 should be attributed to 4C +39.23A and 4C +39.23B. We  examined whether other counterparts make contributions to the $\gamma$-rays of 4FGL J0824.9+3915. We searched for the low-energy counterparts with the SIMBAD Astronomical Database and found 19 sources within the 95\% error circle of 4FGL J0824.9+3915 (covering MJD 57500--58500). We further investigated these sources with the NED. Besides 4C +39.23B and 4C +39.23A, only other two sources (FIRST J082519.5+391646 and SDSS J082507.90+391859.0) are detected by the Wide-field Infrared Survey Explorer (WISE). However, the radio flux at 1.4 GHz of FIRST J082519.5+391646 is much lower than that of 4C +39.23B (more than two orders of magnitude), and no observational data at higher radio frequencies are available, while SDSS J082507.90+391859.0 has no radio observations. Among the four sources, only 4C +39.23A has the X-ray detection. Therefore, the most likely counterparts of the $\gamma$-ray source 4FGL J0824.9+3915 would be 4C +39.23A and 4C +39.23B, although we still cannot conclusively rule out the other sources within the error circle of its position.

\acknowledgments

This work is supported by the National Natural Science Foundation of China (grants 12022305, 11973050, U1738136, U1731239, 11851304, and 12133003), and Guangxi Science Foundation (grants 2017AD22006, 2019AC20334, and 2018GXNSFGA281007).

\clearpage

\clearpage
\begin{deluxetable}{lccccc}
\tabletypesize{\scriptsize} \tablecolumns{6} \tablewidth{0pc}
\tablecaption{Analysis Results of the $\sim$13 yr \emph{Fermi}/LAT Observation Data with the Different Templates}\tablenum{1}
\tablehead{\colhead{Templates}&\colhead{$R.A.$} &  \colhead{$Decl.$}&\colhead{TS}&\colhead{$F_{0.1-300~\rm GeV}$ } &\colhead{$\Gamma_{\rm \gamma}$}\\
\colhead{}&\colhead{[deg]}&\colhead{[deg]}&\colhead{}&\colhead{[10$^{-8}$ photon cm$^{-2}$ s$^{-1}$]}&\colhead{}}
\startdata
4FGL J0824.9+3915&126.236&39.257&492.7&1.60$\pm$0.15&2.48$\pm$0.05\\
\hline
4C +39.23A&126.231&39.278&491.2&1.60$\pm$0.16&2.48$\pm$0.05\\
\hline
4C +39.23B&126.349&39.329&458.6&1.60$\pm$0.16&2.50$\pm$0.05\\
\hline
4C +39.23A&126.231&39.278&316.1&1.30$\pm$0.34&2.48$\pm$0.07\\
 \qquad+\\
4C +39.23B&126.349&39.329&7.8& $<$0.59\tablenotemark{a}&2.52$\pm$0.28\\
\hline
4FGL J0824.9+3915&126.236&39.257&\nodata&\nodata&\nodata\\
\qquad+\\
4C +39.23A&126.231&39.278&\nodata&\nodata&\nodata\\
 \qquad+\\
4C +39.23B&126.349&39.329&\nodata&\nodata&\nodata\\
\enddata
\tablenotetext{a}{An upper-limit of flux is given when TS$<$9. }
\end{deluxetable}

\begin{deluxetable}{lccccccccccccccccccccc}
\tabletypesize{\tiny} \rotate \tablecolumns{22} \tablewidth{0pc}
\tablecaption{SED Fitting Parameters of 4C +39.23B together with that of the other five\tablenotemark{a} CSSs in Zhang et al. (2020)}\tablenum{2}
\tablehead{ \colhead{}  & \colhead{}  &\multicolumn{11}{c}{Compact Core} & \multicolumn{8}{c}{Extended Region} \\
\cline{3-13} \cline{15-22}\\
\colhead{Source} &\colhead{z}  & \colhead{$R$} &
\colhead{$B$} & \colhead{$\delta$}& \colhead{$\Gamma$} & \colhead{$\theta$} & \colhead{$\gamma_{\min}$} & \colhead{$\gamma_{\rm b}$} &\colhead{$\gamma_{\max}$}
&\colhead{$N_{0}$ } & \colhead{$p_1$} & \colhead{$p_2$} &\colhead{}& \colhead{$R$} & \colhead{$B$} & \colhead{$\gamma_{\min}$} & \colhead{$\gamma_{\rm b}$} &\colhead{$\gamma_{\max}$} &\colhead{$N_{0}$ } & \colhead{$p_1$} & \colhead{$p_2$} \\
\colhead{}& \colhead{}  &\colhead{[cm]} & \colhead{[G]} & \colhead{}& \colhead{} & \colhead{[deg]} & \colhead{} & \colhead{} &\colhead{} &\colhead{ [cm$^{-3}$]} & \colhead{} & \colhead{} &\colhead{} & \colhead{[kpc]}& \colhead{[$\mu$G]} & \colhead{} & \colhead{} & \colhead{} & \colhead{[cm$^{-3}$]}& \colhead{} & \colhead{}}
\startdata
4C +39.23B&1.21&1.37E18&0.13&6.5&6.5&8.8&1&2192&1E5&0.13&1.56&4.36&&0.5&860&500&1500&7.5E3&9.0E-02&2.2&4\\
3C 138&0.759&7.44E17&0.6&2.8&5.5&18&1&531&5E5&139.8&1.8&3.46&&2.95&277&5E2&5E3&2.5E5&6E-03&2.2&4\\
3C 216&0.67&1.01E18&0.4&3.6&3.6&15.9&1&1427&1E5&25.3&1.8&4.04&&17.54&66&1E2&2E4&1E6&8.8E-04&2.46&4\\
3C 286&0.849&1.13E18&0.2&4.5&4.5&12.7&1&1666&3E5&5.1&1.6&4.5&&19.18&72&1E3&1.9E4&9.5E5&4.7E-04&2.22&4\\
3C 309.1&0.904&8.75E17&0.22&3.6&2.8&15.5&50&1914&1E5&15&1.5&4.2&&15.65&77&5E2&1.5E4&7.5E5&4.3E-04&2.22&4\\
3C 380&0.692&1.11E18&0.15&4.0&2.4&9.5&1&1845&1E5&40.1&1.68&4.0&&7.11&190&1E2&1.5E4&7.5E5&5E-03&2.4&4\\
\enddata
\tablenotetext{a}{We do not take 4C 15.05 into account here since its type is still debated.}
\end{deluxetable}

\end{document}